\documentclass[a4paper,12pt]{article}

\usepackage{amsmath,amssymb,amsfonts,epsfig,cite,setspace,bigstrut,framed}
\usepackage[all]{xy}
\usepackage{color}
\usepackage{pifont}

\pdfoutput=1

\begin{document}


\vskip 0.25in

\newcommand{\todo}[1]{{\bf ?????!!!! #1 ?????!!!!}\marginpar{$\Longleftarrow$}}
\newcommand{\nn}{\nonumber}
\newcommand{\comment}[1]{}
\newcommand\T{\rule{0pt}{2.6ex}}
\newcommand\B{\rule[-1.2ex]{0pt}{0pt}}

\newcommand{\cM}{{\cal M}}
\newcommand{\cW}{{\cal W}}
\newcommand{\cN}{{\cal N}}
\newcommand{\cH}{{\cal H}}
\newcommand{\cK}{{\cal K}}
\newcommand{\cT}{{\cal T}}
\newcommand{\cZ}{{\cal Z}}
\newcommand{\cO}{{\cal O}}
\newcommand{\cQ}{{\cal Q}}
\newcommand{\cB}{{\cal B}}
\newcommand{\cC}{{\cal C}}
\newcommand{\cD}{{\cal D}}
\newcommand{\cE}{{\cal E}}
\newcommand{\cF}{{\cal F}}
\newcommand{\cX}{{\cal X}}
\newcommand{\IA}{\mathbb{A}}
\newcommand{\IP}{\mathbb{P}}
\newcommand{\IQ}{\mathbb{Q}}
\newcommand{\IH}{\mathbb{H}}
\newcommand{\IR}{\mathbb{R}}
\newcommand{\IC}{\mathbb{C}}
\newcommand{\IF}{\mathbb{F}}
\newcommand{\IV}{\mathbb{V}}
\newcommand{\II}{\mathbb{I}}
\newcommand{\IZ}{\mathbb{Z}}
\newcommand{\re}{{\rm Re}}
\newcommand{\im}{{\rm Im}}
\newcommand{\tr}{\mathop{\rm Tr}}
\newcommand{\ch}{{\rm ch}}
\newcommand{\rk}{{\rm rk}}
\newcommand{\ext}{{\rm Ext}}
\newcommand{\beq}{\begin{equation}}
\newcommand{\eeq}{\end{equation}}
\newcommand{\bea}{\begin{eqnarray}}
\newcommand{\eea}{\end{eqnarray}}
\newcommand{\ba}{\begin{array}}
\newcommand{\ea}{\end{array}}

\newcommand{\CA}{\mathbb A}
\newcommand{\CP}{\mathbb P}
\newcommand{\tmat}[1]{{\tiny \left(\begin{matrix} #1 \end{matrix}\right)}}
\newcommand{\mat}[1]{\left(\begin{matrix} #1 \end{matrix}\right)}
\newcommand{\diff}[2]{\frac{\partial #1}{\partial #2}}
\newcommand{\gen}[1]{\langle #1 \rangle}

\newtheorem{theorem}{\bf THEOREM}
\newtheorem{proposition}{\bf PROPOSITION}
\newtheorem{observation}{\bf OBSERVATION}

\def\theequation{\thesection.\arabic{equation}}
\newcommand{\setall}{
	\setcounter{equation}{0}
}
\renewcommand{\thefootnote}{\fnsymbol{footnote}}

\begin{titlepage} 
  \begin{flushright}
  {\tt\normalsize KIAS-P12047}\\
  \end{flushright}
  \vspace*{\stretch{1}}
  \begin{center}
     \LARGE 
\bf Quiver Structure of Heterotic Moduli
  \end{center}
  \vspace*{\stretch{2}}
  \begin{center}
    \begin{minipage}{\textwidth}
      \begin{center}
        \large         
        Yang-Hui He$^{1}$ and
	Seung-Joo Lee$^2$
      \end{center}
    \end{minipage}
  \end{center}
  \vspace*{1mm}
  \begin{center}
    \begin{minipage}{\textwidth} \it
      \begin{center}
        ${}^1$ School of Physics, NanKai University, Tianjin, 300071, 
	P.R. China\\
        Department of Mathematics, 
       City University, London, EC1V 0HB, U.K.\\
        Merton College, University of Oxford, OX1 4JD, U.K.\\
        ${}^2$ School of Physics, Korea Institute for Advanced Study, Seoul 130-722, Korea
      \end{center}
    \end{minipage}
  \end{center}

  \vspace*{\stretch{1}}
  \begin{abstract}
  We analyse the vector bundle moduli arising from generic heterotic compactifications from the point of view of quiver representations.
  Phenomena such as stability walls, crossing between chambers of supersymmetry, splitting of non-Abelian bundles and dynamic generation of D-terms are succinctly encoded into finite quivers.
  By studying the Poincar\'e polynomial of the quiver moduli space using the Reineke formula, we can learn about such useful concepts as Donaldson-Thomas invariants, instanton transitions and supersymmetry breaking.
  \end{abstract}
  \vspace*{\stretch{5}}
  \begin{minipage}{\textwidth}
    \underline{\hspace{5cm}}
    \centering
    \\
    Email: hey@maths.ox.ac.uk, s.lee@kias.re.kr
  \end{minipage}
\end{titlepage}

\tableofcontents

\newpage

\section{Introduction, Summary, and Prospectus}
The study of moduli constitutes one of the most important subjects in contemporary string theory.
The stabilization of the scalar moduli fields to fix their vacuum expectation values (vevs) to reasonable values is central to any phenomenological model;
in parallel, the space of moduli as an algebraic variety, associated to the geometries which the moduli themselves parametrize, is a concept indispensable to modern mathematics.
Indeed, be it brane-probes on Calabi-Yau singularities, M-theory on manifolds of $G_2$ holonomy, or any other scenario which could lead to desired four-dimensional physics, the structure of the moduli space arising from the geometry and the subsequent generation and extremization of the potential have been under intense investigation.

In the realm of heterotic string compactifications, the oldest approach to string phenomenology, the algebro-geometric nature of the moduli fields is particularly pronounced.
The initial attempts, some three decades ago, of using the tangent bundle of the compacfication Calabi-Yau threefold $X$ to break the $E_8$ gauge theory to an $E_6$ GUT, possessed geometric moduli given by the complex and K\"ahler parametres of the threefold \cite{Candelas:1985en}.
With the advent of more powerful methods in geometry, the more ``general embedding'' of taking stable holomorphic vector bundles $V$ on $X$ with structure group beyond $SU(3)$ and coupled to the presence of Wilson lines (cf.~e.g.~\cite{Buchbinder:2002pr,Donagi:2004qk}) has given us the prospects of realistic MSSM vacua \cite{Braun:2005nv,Bouchard:2005ag,Anderson:2011ns}.
At the same time, this favourable construction produces additional, vector bundle, moduli which need to be addressed.
This in itself has blossomed into a rich subject; q.v.~e.g.~\cite{Curio:2012iv,Costa:2011zzb,He:2003tj,Buchbinder:2002wz,Anguelova:2010qd,Anderson:2010mh,Micu:2009ci,LiQin,LiQin2,FriedmanQin,HL}, of interest to physicists and mathematicians alike.

Our starting point is the Hermitian-Yang-Mills equations for $V$ which guarantees the low-energy theory - whose gauge group is the commutant of the structure group $G$ of $V$ in $E_8$ - to be supersymmetric.
In terms of the connection $A_{\mu}$ on $V$ and the Calabi-Yau metric $g_{\mu \nu}$ of $X$, the equations are
\begin{equation}
F_{ab} = F_{\bar{a} \bar{b}} = 0 \ , 
\quad
g^{a \bar{b}} F_{a \bar{b}} = 0 \ ,
\end{equation}
where $F = dA + A \wedge A$ is the field strength of $A$. 
The first set is a statement of holomorphicity and the second, a set of highly non-trivial partial differential equations.
The moduli space of which we speak is then the space of solutions to these equations.
Physically, the scalar parametres characterizing the solutions obey supersymmetry constraints which can be written as vanishing of D-terms.
Mathematically, the celebrated Donaldson-Uhlenbeck-Yau theorem reduces the solutions to so-called {\it poly-stable} holomorphic bundles and the moduli spaces thereof are intricate algebraic varieties.
The (virtual) dimensions of such varieties are captured by the Donaldson-Thomas  (DT) invariants.

Recently, a systematic outlook was nicely undertaken in \cite{Anderson:2009sw,Anderson:2009nt,Anderson:2010ty} to study the so-called K\"ahler sub-structure of the vector bundle moduli spaces arising from heterotic compactification, whereby providing an algorithmic handle on the geometry.
By identifying {\em stability walls} across which stable bundles become unstable and on which the structure groups ``split'' to lower rank, chambers of supersymmetry-preservation  can thus be mapped.

This decomposition wherein non-Abelian bundles become non-semi-simple, i.e., products of factors of lower rank, and have, in particular, $U(1)$ factors under which the moduli fields may be charged renders the situation especially poignant.
The $U(1)$ groups may be anomalous in the sense of Green-Schwarz and the corresponding D-terms are induced whose Fayet-Iliopoulos (FI) parametres are controlled by the vevs of the moduli fields.
A positive-semi-definite potential is perturbatively generated which becomes positive in chambers of non-supersymmetry and vanishes where supersymmetry is preserved.
Successive application of this decomposition amounts to, mathematically, the usage of the {\it Harder-Narasimhan filtration}.
As a by-product, a recursive algorithm can be established in computing the Donaldson-Thomas invariants.

The astute reader would find the above discussions, on supersymmetry, chambers and stability, reminiscent of another vast subject, that of {\em quiver theories}.
That we have fields charged under product groups with Abelian factors contributing to Fayet-Iliopoulos terms compels us into the territory of quiver representations.
This is of no surprise to us since the extrapolation from the large to the small volume limits in the category of branes, from bundle stability to quiver stability, has been the perspicacious observations of \cite{Douglas:2000ah,Aspinwall:2004mb,Douglas:2000qw}.
The incipience of our analysis will thus be along this train of thought: to reformulate, aided by the string-theoretic language of \cite{Douglas:2000ah,Aspinwall:2004mb,Douglas:2000qw} and the mathematical insights of \cite{king,reineke}, the rich sub-structure of the vector bundle moduli space in heterotic compactifications in terms of quiver theories.

Indeed, we shall see that the quiver perspective on heterotic moduli is the most natural one and is conducive to explicit computation.
The D-terms, charges and interactions amongst the moduli are thus succinctly encoded graphically and with the aid of powerful explicit formulae from the quiver literature, we can visualize the structure of the moduli space and in many cases readily obtain such quantities as the Donaldson-Thomas invariants.
Therefore, our quiver approach to heterotic moduli is not mere linguistic sophistry but of practical value in calculations.

The organization of the paper is as follows.
We begin with detailed description of the two central themes: vector bundle moduli in \S\ref{s:V} and quiver moduli in \S\ref{s:Q}.
We will take care to draw parallels between the two, in the definition of mu- and theta-stability, the wall-separated regions (or chambers) in K\"ahler parametre space wherein stability implies the preservation of supersymmetry, and in the D-terms which encode this information.
Along the way we will discuss the computation of Donaldson-Thomas invariants in \S\ref{introDT} as well as the concept of Pi-stability which extrapolates to the two stabilities in \S\ref{s:pi}.

We then take a unified perspective in \S\ref{s:quiver-bundle} and show how given a heterotic compactification scenario we can draw a quiver, whose moduli space controls the bundle moduli space and, in particular, the phenomena of stability walls and crossings, splittings and calculations of the DT invariants.
Of great use is the explicit formula of Reineke, which we exploit in \S\ref{s:sQV}.
These are supplanted with illustrative examples in \S\ref{s:eg}.
Given the algorithmic power of our approach, we show how one can readily analyze large classes of Calabi-Yau threefolds and demonstrate with a plenitude of concrete examples in \S\ref{s:CICY}.
We will also present some generalities on analytic results in \S\ref{s:ana}, as well as an interesting ``curved'' wall in \S\ref{s:curve}.

The future directions to our quiver-heterotic dictionary are in abundance.
Throughout this paper we have made the assumption that our special unitary bundle $V$ completely splits into line-bundles on the wall, this Abelian split suffices to illustrate whilst significantly simplifies.
Of course, the general situation is to have non-Abelian, sheaf factors; this is an obvious next step.
Furthermore, Reineke's formula is for quivers without loops. To extend this to arbitrary quivers, whereby accommodating non-trivial F-terms as well, is certainly important.

The quiver language and its usage for computation of Donaldson-Thomas invariants also appear in the study of wall-crossing phenomena in $\cN=2$ supersymmetric string theories. 
In type II string theory on Calabi-Yau three-folds, one can think of bound states of elementary D-branes wrapped on various cycles. 
When the central charges of the constituents are nearly aligned, an elegant effective description arises by dimensionally reducing the system to $\cN=4$ quiver quantum mechanics~\cite{Denef:2002ru}. 
In the Higgs branch, the (refined) index is determined by the Poincar\'e polynomial of the moduli space of classical vacua of this quiver. 

Along this vein of thought, the quiver description has recently been used to determine BPS spectrum of a given $\cN=2$ theory~\cite{Alim:2011ae, Alim:2011kw}, and more direct exploration of the cohomology structure of the quiver variety has also been undertaken (mainly) for Abelian quivers (possibly involving loops) in the context of wall-crossing~\cite{Denef:2007vg, Bena:2012hf, Lee:2012sc, Lee:2012na, Manschot:2012rx}. 
Given that the mathematical formulation of the wall-crossing phenomena is very much the same as that of heterotic moduli, it is natural to expect a physical bridge between the two. 
This suggests another interesting direction for further extensions of this work.

\section{Vector Bundle Moduli and $\mu$-Stability}\label{s:V}\setall
Let us begin by introducing the importance of bundle stability in relation to our heterotic moduli.
We recall the standard Donaldson-Uhlenbeck-Yau theorem that a holomorphic vector bundle $V$ admits a connection $A$ solving the Hermitian Yang-Mills (HYM) equations
\begin{equation}\label{HYM}
F_{ab} = F_{\bar{a}\bar{b}} = 0 \ ;
\quad
g^{a \bar{b}} F_{a \bar{b}} = 0 \ ,
\end{equation}
on a Calabi-Yau threefold $X$ with Ricci-flat K\"ahler metric $g_{a\bar{b}}$, if and only if $V$ is {\em poly-stable}.
To define the notion of poly-stability one needs the concept of {\em slope}; this is a quantity $\mu_B$, which for any coherent sheaf $\cF$ is defined as
\begin{equation}\label{muB}
\mu_B(\cF) := \frac{1}{\rk(\cF)} \int_X c_1(\cF) \wedge \omega \wedge \omega 
= \frac{1}{\rk(\cF)} d_{ijk} c_1(\cF)^i t^j t^k \ .
\end{equation}
Here, we have explicitly placed a subscript ``B'' to emphasize that this $\mu$-slope is for bundles, in order to differentiate from a similar slope which we will later define for quivers.
Furthermore, $\omega = t^i \omega_i$ is the K\"ahler form expanded in some basis $\omega_i$ of $H^{1,1}(X; \IR)$ and $d_{ijk} = \int\limits_X \omega_i \wedge \omega_j \wedge \omega_k$ are the triple-intersection numbers.

A vector bundle $V$ is {\em stable} if all proper subsheaves $\cF \subset V$ with rank $0 < \rk(\cF) < \rk(V)$ obey the inequality
\begin{equation}\label{stab}
\mu_B(\cF) < \mu_B(V) \ .
\end{equation}
It is poly-stable if it is a direct sum of bundles each of which is stable and all of which have the same slope with respect to a fixed choice of $\omega$ called a {\bf polarization}.
Furthermore, if the strict inequality \eqref{stab} is relaxed to $\mu_B(\cF) \le \mu_B(V)$, then $V$ is called semi-stable.
However, we will henceforth only consider stable and poly-stable sheaves because these match one-to-one with the solutions to HYM equations.
Indeed, when we discuss Donaldson-Thomas invariants in \S\ref{introDT}, these are only defined when the moduli spaces of semi-stable and stable sheaves coincide \cite{Joyce}.
The criterion \eqref{muB} for determining stability is sometimes referred to as the Gieseker-Mumford-Takemoto {\bf $\mu$-stability}.

In the poly-stable case, we have that
\begin{equation}\label{Vsum}
V = \bigoplus_{i=1}^m V_i \ , \mbox{ with } \mu_B(V) = \mu_B(V_i) \ ,
\end{equation}
where the direct (Whitney) summands $V_i$ are all of the same slope.
Note that the division by the rank in the definition of slope exhibits its importance here: the additivity of the Chern character implies that for the form in  Eq.~\eqref{Vsum}, $\rk(V) = \sum\limits_{i=1}^m \rk(V_i)$ and 
$c_1(V) = \sum\limits_{i=1}^m c_1(V_i)$. 
Thus, $\mu_B(V) = \mu_B(V_i)$ is a possible solution under these two constraints.

Because the solutions to the HYM equations are in one-one correspondence with poly-stable bundles and stability explicitly depends on the choice of K\"ahler moduli $t^i$, there exist {\em stability walls} in K\"ahler moduli space demarcating regions of where the HYM equations are obeyed.
These walls thus divide the K\"ahler moduli space into {\bf chambers} wherein supersymmetry is preserved or broken.
In the chambers of supersymmetric K\"ahler moduli, there are, in addition, the (stable) vector bundle moduli to which we shall shortly turn.
In particular, the dichotomy between poly-stable and stable is an important one; it reflects the structure of the bundle as either being a trivial direct sum or not.

\subsection{Split Extensions and Poly-stability}
Suppose that a vector bundle $V$ had a subsheaf $\cF$, then we have the short exact sequence
\begin{equation}\label{seq}
0 \rightarrow \cF \rightarrow V \rightarrow Q \rightarrow 0 \ .
\end{equation}
Here, $\cF$ injects into $V$ (whence subsheaf) and $Q$ can be thought of as the quotient $V / \cF$.
Conversely, $V$ can be regarded as an {\em extension} of $Q$ by $\cF$.
The moduli space of such possible extensions is measured by the first (sheaf) Ext group \cite{hart}
\begin{equation}\label{ext1}
\mbox{Ext}^1(Q,\cF) \simeq H^1(X, \cF \otimes Q^*) \ ,
\end{equation}
where $Q^*$ is the sheaf dual.
The trivial element in this space of extensions is, of course, the case where $V$ splits simply into the direct sum: $V \simeq \cF \oplus Q$.

The subsheaf $\cF$ {\em de-stabilizes} $V$ (whereby making it unstable) if $\mu_B(\cF) \ge \mu_B(V)$ whereas $V$ is stable if there exists no such $\cF$.
Hence, in K\"ahler moduli space, the walls on which regions of stability meet instability must have $\mu_B(\cF) = \mu_B(V)$ for some of its subsheaves.
These are precisely the circumstances where $V$ is poly-stable wherein $V$ is a direct sum of sub-bundles of the same slope as in Eq.~\eqref{Vsum}.
We summarize by saying that (cf.~section 2 of \cite{Anderson:2010ty})
\begin{proposition}\label{split}
On the walls of stability in K\"ahler moduli space, the vector bundle splits into a direct sum of sub-bundles of equal $\mu$-slope.
\end{proposition}

One could iterate this procedure of quotienting by sub-sheaves and essentially break up the vector bundle $V$ into an extension of sub-sheaves. We will soon take these to be the simplest constituents, viz.,~line bundles.
This is a celebrated result \cite{HN} (cf.~more pedagogical details in \cite{HL}) which states that 
\begin{theorem}\label{filter}
[Harder-Narasimhan]
For a holomorphic vector bundle $V$ on a closed K\"ahler manifold $X$, 
\begin{itemize}
\item There exists a filtration
\begin{equation}
0 = \cF_0 \subset \cF_1 \subset \ldots \subset \cF_m = V
\end{equation}
such that $\cF_i /  \cF_{i-1}$ are semi-stable sheaves for $i=1, \ldots, m$ with $\mu_B(\cF_{i}/\cF_{i-1}) > \mu_B(\cF_{i+1}/\cF_{i})$;

\item If $V$ is semi-stable, then  $\cF_i /  \cF_{i-1}$ are stable with  
$\mu_B(\cF_{i}/\cF_{i-1}) = \mu_B(V)$ for all $i$;

\item The graded sum $\cF_1 \oplus \cF_2/\cF_1 \oplus \ldots \oplus \cF_m/\cF_{m-1}$ is uniquely determined up to isomorphism.
\end{itemize}
\end{theorem}
This filtration thus gives us a whole list of intertwined short exact sequences of the form \eqref{seq}:
\begin{equation}\label{filter2}
\begin{array}{c}
0 \rightarrow \cF_1 \rightarrow \cF_2 \rightarrow \cF_2 / \cF_1 \rightarrow 0 \ , \\
0 \rightarrow \cF_2 \rightarrow \cF_3 \rightarrow \cF_3 / \cF_2 \rightarrow 0 \ , \\
\vdots\\
0 \rightarrow \cF_{m-1} \rightarrow V \rightarrow V / \cF_{m-1} \rightarrow 0 \ .
\end{array}
\end{equation}

The physical realization of the Harder-Narasimhan filtration as splitting on the stability walls is nicely described in \cite{Anderson:2009nt}.
Combining Proposition \ref{split} and Theorem \ref{filter}, we have, in general, that a rank $n$ bundle $V$ can decompose into a direct sum of lower rank sheaves on the stability wall.
Henceforth - having in mind phenomenological applications - we will only consider $V$ with {\it special unitary} structure group, hence, $c_1(V)$ and consequently $\mu_B(V)$ both vanish.
Therefore, marking the ranks of bundles as superscripts for clarity, we have 
\begin{equation}\label{Vsplit}
V^{(n)} \longrightarrow \bigoplus_{i=1}^m V_i^{(n_i)} \ ,
\qquad
\mu_B(V_i) = \mu_B(V) = 0 \ .
\end{equation}

Of the $m$ terms in the decomposition, there will be up to $m-1$ anomalous $U(1)$ gauge factors. 
We can see this from the simple decomposition of $SU(n)$
\beq
SU(n) \to S[U(n_1) \times \cdots \times U(n_m)] \simeq SU(n_1) \times \cdots SU(n_m) \times U(1)^{m-1} \ , 
\eeq
at the Lie algebra level. 
In the low energy, to each of the $U(1)$ factors there will be an associated D-term, with a Fayet-Iliopoulos (FI) parametre dependent upon K\"ahler moduli.
We will return to this point in \S\ref{Dterm}.

Because each such term in the extensions \eqref{filter2} has associated moduli computed by \eqref{ext1}, we will have a total of $m^2 -m$ different types of ``bi-fundamental'' bundle moduli given by 
\begin{equation}\label{ext-moduli}
\mbox{Ext}^1(V_j, V_i) = H^1(X, V_i \otimes V_j^*) \ , \qquad i \ne j = 1, \ldots, m \ .
\end{equation}
When some of the $V_i$ are non-Abelian, we also have adjoint bundle moduli in ${\rm Ext}^1(V_i , V_i) = H^1(X,V_i \otimes V_i^*)$.
However, in this paper we only consider complete splits into line bundle summands and hence, the adjoint moduli do not appear.


\subsection{Vector Bundle Moduli and Donaldson-Thomas Invariants}\label{introDT}
Finding and proving the stability of vector bundles, even though being a reduction of the complicated Hermitian Yang-Mills equations as a set of partial differential equations to an algebraic formulation, are still rather difficult.
Nevertheless, one could enumerate them without explicitly finding a solution.
This enumeration of stable bundles $V$ of a fixed total Chern class
\begin{equation}
\bold{c}(V) = (\rk, c_1, c_2, c_3)
\end{equation}
with respect to some K\"ahler polarization $\omega$ is provided by the powerful Donaldson-Thomas invariants.
The subject is a vast one and the reader is referred to, for example, Refs.~\cite{DTGW,Joyce}.

First let us define the moduli space, for a Calabi-Yau threefold $X$,
\begin{equation}\label{Mbundle}
\cM_V(X, \bold{c}; \omega) = \{
\mbox{stable sheaves on $X$ with Chern class } \bold{c}, 
\mbox{ $\mu_B$-stable with respect to $\omega$}
\} \ ;
\end{equation}
that this moduli space is well-defined is a subtle issue \cite{DT,Thomas}, allowing for the definition of a {\em virtual fundamental class} $[\cM_V(X,\bold{c}; \omega)]^{\mbox{{\tiny Vir}}}$ of zero degree.
The integral over this fundamental class is essentially the ``volume'' of the moduli space which constitutes a {\em count} of the number of stable sheaves; it was shown in Refs.~\cite{DT,Thomas} that the result is indeed an integer, now called the Donaldson-Thomas (DT) invariant:
\begin{equation}
DT(X,\bold{c}; \omega) = \int_{[\cM_V(X,\bold{c}; \omega)]^{\mbox{{\tiny Vir}}}} 1 \in \IZ \ .
\end{equation}

Subsequently, it was shown in Ref.~\cite{behrend} that when the moduli space is smooth, the virtual fundamental class simply becomes the top Chern class, whence reducing the DT invariant to a signed Euler number:
\begin{equation}
DT(X,\bold{c}; \omega) = \int_{[\cM_V(X,\bold{c}; \omega)]} c_{top}(T^*\cM_V(X,\bold{c}; \omega))
= (-1)^{\dim(\cM_V(X,\bold{c}; \omega))}\chi(\cM_V(X,\bold{c}; \omega)) \ .
\end{equation}
This fact will be of greatest computational use to us.
Note that this is a signed invariant depending on whether the dimension of the moduli space is even or odd.

In passing, we also mention that the generating function for the DT invariants (cf.~\cite{balazs1,balazs2,LiQin}) often takes remarkable forms in terms of well-known combinatorial functions.
For example, on the complete intersection Calabi-Yau threefold $Y = [\IP^5 | 2, 4]$ which is the generic intersection of a quadric and a quartic in $\IP^5$, the generating function for stable sheaves of Chern class $\bold{c}=({\rm rk}, 2, 0, -m)$ is
\begin{equation}
\sum\limits_{m \in \IZ} DT(Y, \bold{c} = ({\rm rk}, 2, 0, -m); \omega) q^m = 
2 M(q^2)^{2 \chi(Y)} \ ;
\end{equation}
where $\chi(Y) = -176$ and $M(q) = \prod\limits_{n=1}^\infty (1 - q^n)^{-n}$ is the famous McMahon function.

\section{Quiver Moduli Space and $\theta$-Stability}\label{s:Q}\setall
Having expounded on the requisites from the bundle side,
in this section, we give some rudiments of the representation theory of quivers, emphasizing the issue of stability in a context parallel to the above.
The interested reader is referred to, for example, Ref.~\cite{Derksen} for a more in depth presentation of this material.

A quiver diagram is defined as a pair $\cQ= (\cQ_0, \cQ_1)$ where $\cQ_0$ is a finite set of vertices (or nodes) and $\cQ_1$ is a finite set of oriented edges connecting these vertices. 
For $\rho\in \cQ_1$ we let $h(\rho)$ to denote the vertex attached to the head of the arrow and $t(\rho)$ the one to the tail. 
A path in $\cQ_1$ is a sequence $x=\rho_1\ldots\rho_n$ of arrows such that $h(\rho_{i+1})=t(\rho_i)$. Moreover, for each vertex $v\in\cQ_0$ we could consider a trivial path $e_v$ which starts and ends in $v$. The {\em path algebra} $k\cQ$ associated with the quiver is the $k$-algebra whose basis is the collection of paths with the product rule given by concatenation of the paths, and $k$ is some ground number field, usually taken to be $\mathbb{C}$. 
That is, the multiplication is $x\cdot y = xy$ if $h(y)=t(x)$ and 0 otherwise.

A {\bf representation} of a quiver is the assignment of a vector space $V_v$ to each vertex $v\in\cQ_0$ and a linear map $f_\rho:\;V_{t{(\rho)}}\mapsto V_{h{(\rho)}}$ to each arrow $\rho\in\cQ_1$.
We can collect the dimensions of the various vector spaces into a {\em dimension vector} $\bold d=(\dim V_1,\ldots,\dim V_m)\in\mathbb{Z}^{m}$
where $m=|\cQ_0|$ is the number of vertices.
Different representations of a given quiver are different sets of vector spaces (respectively, morphisms) that one can assign to each vertex (respectively, edge).

We can define a homomorphism between two representations $R$ and $S$ of the same quiver, as a set of linear maps $\phi_v : V_{R,v} \mapsto V_{S,v}$ for each vertex $v \in \cQ_0$ satisfying $f_{R,\rho} \phi_{h(\rho)}  = \phi_{t(\rho)} f_{S,\rho}$ where composition of maps is from left to right.
That is, we have the commutative diagram
\begin{equation}
\begin{xy}
(0,20)*+{V_{R,t(\rho)}} = "a"; (50,20)*+{V_{R, h(\rho)}} = "b"; 
(0, 0)*+{V_{S,t(\rho)}} = "c"; (50, 0)*+{V_{S,h(\rho)}} = "d"; 
(100,20)*+{\mbox{Representation $R$}};
(100, 0)*+{\mbox{Representation $S$}};
{\ar@{->}^{f_{R,\rho}} "a"; "b"};
{\ar@{->}_{f_{S,\rho}} "c"; "d"};
{\ar@{->}_{\phi_{t(\rho)}} "a"; "c"};
{\ar@{->}^{\phi_{h(\rho)}} "b"; "d"};
\end{xy}
\end{equation} 
If $\phi$ is an injective homomorphism into $S$ then $R$ is a sub-representation of $S$.

Now we are ready to define a notion of stability for the representation of quivers, as introduced by King in Ref.~\cite{king}; this is the so-called {\bf $\theta$-stability}.
As in the case of $\mu$-stability, this is defined with respect to some choice of parametres in analogy to the $t^i$ in Eq.~\eqref{muB}, customarily denoted as $\theta_v$ (whence the name).
We learn from Ref.~\cite{king} that for a representation $S$ with dimension vector $\bold d({S}) = (d_1(S), \ldots, d_{m = |\cQ_0|}(S))$, if we could find $\theta_v \in \IZ$ such that
$
\sum\limits_{v=1}^m \theta_v d_v(S) = 0
$ 
and 
$
\sum\limits_{v=1}^m \theta_v d_v(R) > 0
$
for {\em any} proper sub-representation $R$ of $S$, then $S$ is $\theta$-stable.
Noting the greater-than sign which had been originally introduced into the literature, it will be more convenient to adhere to the notation of \cite{reineke} and to define a $\mu$-slope which parallels 
\eqref{muB}. 
For a dimension vector $\bold d(R)$ corresponding to representation $R$, let
\begin{equation}\label{muQ}
\mu_Q(\bold d(R)) := \frac{\sum\limits_{v=1}^m \theta_v d_v}{ \sum\limits_{v=1}^m d_v }
\qquad
\mbox{for }  \theta_v \in \IZ \ .
\end{equation}
Note that we have inserted the subscript ``Q'' to emphasize that this $\mu$-slope is for quivers, in contrast to $\mu_B$.
The $\theta$ parametres are the analogues of the  K\"ahler polarization $\omega$ in the vector bundle case.
Thus defined, a representation $S$ is stable if for {\em every} proper sub-representation $R$, we have
\begin{equation}\label{Qstab}
\mu_Q(R) < \mu_Q(S) \ ,
\end{equation}
for some fixed choice of polarization $\theta_v$.
We note that this choice of $\theta$ will be negative of the convention chosen in Ref.~\cite{king} in light of the above discussions. 

\subsection{Quiver Moduli Space}
Using geometric invariant theory (GIT), Ref.~\cite{king} constructed the notion of moduli space of representations: this is the quotient of the total vector space of maps by the symmetry group of conjugations.
Specifically, given the representation $(V_v, f_\rho)$ for the quiver as defined above, clearly $f_\rho \in \hom(V_{t(\rho)},V_{h(\rho)})$ and there is a symmetry $g \in GL(V_v)$ acting by conjugation as $(g \cdot f)_\rho := g_{h(\rho)} f_\rho  g_{t(\rho)}^{-1}$. Therefore we define the moduli space as
\begin{equation}\label{Mquiver}
\cM_\cQ(\bold d) = \bigoplus\limits_{\rho \in \cQ_1} \hom(V_{t(\rho)},V_{h(\rho)}) \slash
    \prod_{v \in \cQ_0} GL(V_v) \ ,
\end{equation}
where $\bold d=(\dim V_1,\ldots,\dim V_m)$ is the dimension vector for the representations.
Now, for $\theta$-stable representations, Mumford's method of GIT shows that this is a so-called ``fine'' moduli space \cite{king}, which is the case we consider here.
In such cases, the moduli space is well-defined and is itself a projective variety.

The vigilant reader will have noticed that we have not mentioned anything about relations. Indeed, to the path-algebra introduced above, we can impose formal algebraic constraints amongst the maps $f_\rho$, which in general will be of the form $\{P_j(\{f_\rho\}) = 0\}$ for some polynomials $P_j$ in terms of the arrows.
Such a quiver is called quiver with {\bf relations}.
Subsequently, in the definition \eqref{Mquiver} of the moduli space, we must consider not only the quotient by the symmetry group, but also by the constraints imposed by the relations in the path algebra.

The particular case of the $P_j$ polynomials being the Jacobian of a single function $W(\{f_\rho\})$ is of special interest:
\begin{equation}\label{fterm}
P_j(\{ f_\rho \}) = \frac{\partial W(\{f_\rho\})}{ \partial f_j} = 0 \ .
\end{equation}
In supersymmetric gauge theories whose bi-fundamental and adjoint matter contents can be encoded into a quiver representation, and whose superpotential\footnote{This superpotential should not be confused with that coming from the bundle moduli side.} is some polynomial $W$ in terms of the fields, Eq.~\eqref{fterm} precisely prescribes the F-term relations coming from $W$.
Of course, the superpotential $W$ itself must come from gauge invariant terms corresponding to cycles or loops in the quiver.
Because our primary concern, for the sake of simplicity, will be quivers without loops, 
we shall ignore the relations in this paper. 

\subsection{$\Pi$-Stability}\label{s:pi}
Having introduced two moduli spaces and two parallel notions of stability, one from algebraic geometry and the other, from representation theory, it is natural to wonder about their connexion.
Indeed, the purpose of much of the ensuing investigations will be to explicitly use the techniques of one to address the other.
The two concepts of stability have been well established \cite{Douglas:2000ah,Aspinwall:2004mb, Douglas:2000qw} to be related and bring us to the subject of $\Pi$-stability, which is a string-thereotic construct extrapolating to $\mu$- (respectively $\theta$-) stability in the large (respectively small) volume limit of compactification.

Following Refs.~\cite{Douglas:2000ah, Douglas:2000qw}, for a holomorphic cycle in the Calabi-Yau manifold $X$, one could consider it as a support for some sheaf, call it $E$, such as in the situation where $E$ is a supersymmetric cycle wrapped by a brane whose world-volume theory is a gauge theory with connection on the supported sheaf. The central charge of the brane can then be defined as
\begin{equation}
Z = \ch(E) \cdot \Pi \ ,
\end{equation}
where $\mbox{ch}(E)$ is the Chern character of $E$ and $\Pi$ is the vector of periods, which consists of integrals of powers of the K\"ahler form over appropriate even cycles (equivalently, this is  
$\int_C \Omega$ of the holomorphic 3-form $\Omega$ over 3-cycles in the mirror Calabi-Yau manifold).
In other words, $Z = \sum\limits_{a=0}^3 \ch_a(E) \int_{[C^{(2a)}]} J^a$.
Explicitly, in terms of the triple intersections $d_{ijk} = \int_X \omega_i \wedge \omega_j \wedge \omega_k$ and for the K\"ahler form $J = T^i\omega_i $ where $T^i = B^i + i V^i$ are the complexified K\"ahler parametres, the periods are simply
\begin{equation}
\{
\Pi_0, \Pi^i_2, \Pi^i_4, \Pi_6
\}
=
\{
-1 \ , \quad T^i\ , \quad- \frac12 d_{ijk} T^j T^k \ , \quad 
	\frac{1}{6} d_{ijk} T^i T^j T^k
\} \ .
\end{equation}

One can then define 
\begin{equation}
\phi(E; \Pi) = \frac{1}{\pi} \arg Z(E; \Pi) = \frac{1}{\pi} \im \log \left( 
\sum\limits_{a=0}^3 \ch_a(E) \int_{[C^{(2a)}]} J^a \right) \ ,
\end{equation}
dependent on the choice of complex structure through the period.
Thus, as before we call $E$ {\bf $\Pi$-stable} if every sub-bundle $E'$ has $\phi(E';\Pi) < \phi(E; \Pi)$  for some chosen period $\Pi$ (whence the name).

In the large volume limit where $V \gg 1$, the $\Pi$-slope reduces to
\begin{eqnarray}
\nn
\phi(E; \Pi) &=& \frac{1}{\pi} \im \log \Pi_6 + 
	\frac{1}{2\pi} \im \frac{1}{\Pi_6 \rk(E)} \int_X J \wedge J \wedge c_1(E) \\
&\simeq & \frac32 + \frac{3}{\pi V} \frac{1}{\rk(E)}  \int_X J \wedge J \wedge c_1(E) +
	\cO(V^{-2}) \ .
\end{eqnarray}
This is precisely the notion of $\mu$-stability in Eq.~\eqref{muB}.

On the other hand, near the orbifold limit where the Calabi-Yau can be locally modeled by the affine orbifold variety $\IC^3 / \Gamma$ for some discrete finite subgroup $\Gamma \subset SU(3)$, the brane world-volume theory is a quiver gauge theory \cite{Douglas:1996sw,Lawrence:1998ja,Hanany:1998sd}.
Here, we have D-term contributions to the effective potential as $\sum\limits_v d_v (\theta_v - \zeta_v)^2$ where $\zeta_v$ are FI parametres.
Thus, one can define
\begin{equation}\label{theta-zeta}
\theta_v = \zeta_v - \frac{\sum\limits_{w} \zeta_w d_w}{\sum\limits_{w} e_w d_w} e_v \ , \quad
\zeta_v := \im \Pi_v
\end{equation}
for $\bold e$ being the vector with all entries 1 and $\bold d$ the dimension vector of the representation as previously defined\footnote{
Note that because we have reversed the sign of $\theta$ in the definition of $\mu_Q$ in analogy to $\mu_B$, the sign of the FI parametres is chosen to be positive, as opposed to \cite{Douglas:2000ah,Aspinwall:2004mb, Douglas:2000qw}.
}.
Indeed, substituting this expression into \eqref{muQ} and the stability condition \eqref{Qstab}, we have that a representation $S$ with dimension vector $\bold d(S)$ is stable if for every proper sub-representation $R$, 
\begin{equation}
\frac{\sum\limits_{v} \zeta_v d_v(R)}{\sum\limits_{v} d_v(R)}
<
\frac{\sum\limits_{v} \zeta_v d_v(S)}{\sum\limits_{v} d_v(S)} \ .
\end{equation}
Thus we rephrase $\theta$-stability in terms of a constraint on the FI parametres.

\section{Quiver Representation of Bundle Moduli}\label{s:quiver-bundle}\setall
Given our two parallel skeins of development, especially in the concepts of stability which can be extrapolated as two different limits, it is natural to enquire whether a stronger tie exists.
Our starting point is that Eq.~\eqref{ext-moduli} should be reminiscent of the description of quiver gauge theories using exceptional collections of sheaves, particularly when computing the bi-fundamentals and Yukawa couplings for D-brane probes over local Calabi-Yau singularities \cite{Wijnholt:2002qz,Herzog:2003zc,Herzog:2003dj,Feng:2002kk}.

More precisely, given a split of the form \eqref{Vsplit}, we can associate a node to each direct summand $V^{(n_i)}_i$ and a number of arrows from node $i$ to $j$ dictated by the dimension of the group $\ext^1(V_i, V_j)$.
This is now a finite quiver with representation $(V^{(n_i)}_i, \ext^1(V_i, V_j))$.
We point out that in the quiver literature, especially for the exceptional collections for brane-probe theories \cite{Wijnholt:2002qz,Herzog:2003zc,Herzog:2003dj,Feng:2002kk}, one is more accustomed to the arrows in the graph corresponding to $\ext^0$ since these are $\hom$-maps and thus prescribe natural morphisms in the quiver representation.
Here, because the bundle moduli are given by the first cohomology, our arrows in the quiver will be associated with $\ext^1$.

For simplicity, in this paper, {\em we will restrict to the following assumptions}, which will be seen to be sufficient in giving us a rich and illustrative structure:
\begin{enumerate}
\item complete splitting into line bundles so that each $V^{(n_i)}_i$ is some line bundle $L_i$ and hence all ranks $n_i = 1$. Thus, $V^{(n)} = \bigoplus\limits_{i=1}^n L_i$;
\item there are no loops (directed closed paths or self-adjoining arrows) in the quiver.
\end{enumerate}
It is clearly an interesting immediate future direction to relax these two artificial constraints, a task we will leave to forth-coming work.

A typical portion of our quiver will thus look like
\begin{equation}
\begin{xy}
(0, 0)*+{\stackrel{L_j}{\bullet}} = "c"; (25, 0)*+{\stackrel{L_i}\bullet} = "d"; 
{\ar@{->}^{l_{ij}} "c"; "d"};
\end{xy}
\end{equation}
where we have denoted the dimension of the corresponding portion of the bundle moduli by 
\begin{equation}\label{adj}
l_{ij} \equiv {\rm dim}\;  \ext^1(L_j, L_i) =  h^1 (X, L_i \otimes L_j^*) \ .
\end{equation}
In other words, $l_{ij}$ is the {\bf adjacency matrix} of a loop-less quiver; in particular, its diagonal vanishes.
Note that this is transpose of the usual definition of adjacency matrix. 
This, in a sense, is more natural due to the cohomology structure in Eq.~\eqref{adj}.

Therefore, in this way, we have given a quiver structure to the computation of bundle moduli, in the spirit of Refs.~\cite{reineke,quiverDT, quiverRev}.
This should allow us to use powerful techniques from quiver theory as a handle on the rather complicated object of vector bundle moduli space, and whence moduli arising from heterotic compactifications.

\subsection{Two Perspectives on D-Terms}\label{Dterm}
It has indeed been shown that when the bundle is decomposed completely into Abelian pieces, the bundle moduli on the wall are subject to the D-term and F-term constraints~\cite{Sharpe:1998zu, Anderson:2009nt, Anderson:2009sw, Anderson:2010mh}; a crucial observation is that the stability and holomorphicity parts in the HYM equations, Eq.~\eqref{HYM}, correspond respectively to the D-terms and F-terms. 
Thus, at least locally, the quiver moduli space should be a correct description for the moduli space of holomorphic and stable sheaves. 

Now, the F-terms - essentially the first pair of HYM in the bundle context and 
the contribution from superpotential terms in the quiver context - are a priori constrained, the more non-trivial object of our present concern comes from the D-terms.
To these we now briefly turn.

Suppose our bundle $V$ splits to have a $U(1)$ factor along some stability wall and $\cF$ is the destabilizing sub-sheaf of $V$.
Then, the moduli fields $\phi^i$ will have charge $Q^i$ under this $U(1)$~\cite{Anderson:2009nt} and the corresponding D-term is
\begin{equation}\label{bundleD}
D(\phi^i, t^a) = f(t^a) - \sum_{i,\bar{j}} Q^i G_{i\bar{j}} \phi^i \bar{\phi}^{\bar{j}} \ ,
\quad
f(t^a) \sim \mu_B(\cF) / V \ .
\end{equation}
In the above, $t^a$ are our usual K\"ahler parametres, $G_{i\bar{j}}$ is some positive definite metric, $V$ is the volume of $X$, and the nice fact is that the FI-parametre $f(t^a)$ is, up to normalization, exactly the mu-slope of the sub-bundle $\cF$.
In the simplest case where all the moduli fields $\phi$ are negatively charged, $Q^i <0$.
Then, on the wall $\mu_B(\cF) = 0$ and the vevs of $\phi^i$ are forced to vanish, in the region of stability $\mu_B(\cF) < 0$ and the vevs adjust to compensate, and in the region of instability $\mu_B(\cF) >0 $ and hence supersymmetry is broken by D-terms.

From the quiver perspective, the fact that Abelian group factors give rise to non-zero FI parametres is reflected by the incidence information of the graph.
We recall that (note we use the opposite sign from Ref.~\cite{Feng:2000mi}) the {\bf incidence matrix} $\iota_{v i}$ of the quiver is defined to be $n \times k$, where $n$ is the number of nodes and $k$, the number of arrows such that
\begin{equation}\label{inc}
\iota_{vi} = \left\{
\begin{array}{rcl}
+1 & \mbox{ if } & \mbox{arrow $i$ has tail at $v$} \\
-1 & \mbox{ if } & \mbox{arrow $i$ has head at $v$} \\
0  & \mbox{ otherwise } \ .
\end{array}
\right.
\end{equation}
For loop-less quivers this matrix has the same information as the adjacency matrix in \eqref{adj}.
Subsequently, the $v$-th D-term corresponding to the $v$-th $U(1)$ gauge group factor is simply
\begin{equation}\label{quiverD}
D_v = \zeta_v - \sum\limits_{i=1}^k \iota_{v i} |\phi^{(i)}|^2  \ , \qquad
v = 1, 2, \ldots, n \ ,
\end{equation}
where $\zeta_v$ are FI parametres dependent on K\"ahler moduli.
The encoding of the charges of the moduli fields by $\pm 1$ is a consequence of the Abelian nature of our quivers; this ensures that the moduli space is a toric variety as dictated by the gauged linear-sigma model description.
We shall bear \eqref{bundleD} and \eqref{quiverD} in mind as we proceed.

\subsection{Stable Quivers for Stable Bundles}\label{s:sQV}
An immediate consequence of our identification is that we can obtain geometrical information for $\cM_V(X, \bold{c}; \omega)$ in Eq.~\eqref{Mbundle}.
In Ref.~\cite{reineke}, Reineke gave a beautiful explicit formula (cf.~section 5 of Ref.~\cite{Denef:2002ru}) for the Betti numbers of the moduli space of (semi-)stable quiver representations for a quiver without loops, given an arbitrary dimension vector $\bold d$ and $\theta$-parametres.
As pointed out in Refs.~\cite{reineke, Denef:2002ru}, the moduli spaces of semi-stable and stable representations are the same when $d_v$ are coprime and $\theta_v$ are linearly independent over $\IQ$, except for the trivial relation $\sum_v d_v \theta_v = 0$. 
We only restrict to these cases and hence, we will use the two terms interchangeably.
Indeed, this parallels our above discussions on DT invariants.

The formula is given in terms of the generating function, viz., the Poincar\'e polynomial, of the Betti numbers $b^i$ of the quiver moduli space $\cM_\cQ(\bold d; \boldsymbol{\theta})$ as a projective variety: 
\begin{equation}
P(t) = \sum\limits_{i = 0}^{\dim (\cM_\cQ)} b^i t^i \ .
\end{equation}
In particular, evaluated at $-1$, $P(-1)$ simply gives the Euler characteristic \footnote{For a recent study of Poincar\'e polynomials for Calabi-Yau geometries, q.v.~\cite{Ashmore:2011yw}.}.
The result is as follows.
\begin{theorem}\label{Pt} [Reineke]
Given a loop-less quiver $(\cQ_0, \cQ_1)$ with representation $(V_{v \in \cQ_0}, f_{\rho \in \cQ_1})$ and dimension vector $d_v = \dim V_v$, the Poincar\'e polynomial of the moduli space $\cM_\cQ(\bold d; {\boldsymbol \theta})$ of $\theta$-semistable representations is
\[
P(t) = (t^2-1)^{1 - \sum\limits_v d_v} \
	t^{-\sum\limits_v d_v(d_v-1) } \
	\sum\limits_{\bold d^*} (-1)^{s-1} t^{2 \sum\limits_{k \le l}
		\sum\limits_{v \rightarrow w} d_v^l d_w^k} \
	\prod\limits_{k,v} \left( [ d_v^k]_{t^2}! \right)^{-1} \ .
\]
\end{theorem}
Here, with the convention that $[0]_q = [0]_q! = 1$,
\begin{equation}
[N]_{q} := \frac{1-q^N}{1-q}
\end{equation}
is the standard $q$-bracket and
\begin{equation}
[N]_q! := \prod\limits_{k=1}^N [k]_q = \frac{1-q}{1-q} \frac{1-q^2}{1-q}
\ldots \frac{1-q^N}{1-q}
\end{equation}
is the $q$-factorial function.
The $\bold d^*$ sum is over all ordered partitions of the dimension vector $\bold d$ by non-zero vectors (with non-negative entries) $\bold d^* = (\bold d^1, \ldots, \bold d^s)$ so that $d_v = \sum\limits_{k=1}^s d_v^k$ and the $\theta$-stability conditions $\sum\limits_v \theta_v (\sum\limits_{l=1}^k d_v^l)> 0$ are satisfied for all $k = 1, \ldots, s-1$.

Now, as explained above, we are only considering complete splits into line bundles, so our dimension vector is simply a list of 1's.
In this case, all $d_v^k$ are equal to 0 or 1 and the $q$-factorial values are all 1; hence,  Theorem \ref{Pt} simplifies to 
\begin{equation}\label{PtAb}
P(t) = (t^2 -1)^{1-n} \ 	
\sum\limits_{\bold d^*} (-1)^{s-1} t^{\ 2 \sum\limits_{k \le l}
		\sum\limits_{v \rightarrow w} d_v^l d_w^k}  \ .
\end{equation}
The reader may find the indices in the above equations overwhelming; we will be very explicit in our illustrative examples below.

\subsection{Two Illustrative Examples}\label{s:eg}
Let us now illustrate some examples of quiver structure and present how it can be used to systematically determine bundle moduli.
The examples given in this subsection were already analyzed in the context of DT invariants in section 5 of Ref.~\cite{Anderson:2010ty}, following Refs.~\cite{LiQin, LiQin2}. Here we shall mainly focus on how the same results are reproduced in our language.

\subsubsection{A Rank-2 Example: $SU(2) \to S[U(1) \times U(1)]$}
Let us take 
\beq\label{X24}
X=\left[
\begin{array}
[c]{c}%
\IP^1\\
\IP^3
\end{array}
\left|
\begin{array}
[c]{c}%
2\\
4 
\end{array}
\right.  \right] \ ,
\eeq
as a bi-degree $(2,4)$ hypersurface in $\IP^1 \times \IP^3$ and consider $SU(2)$ bundles on $X$ with the fixed total Chern class
\beq\label{c}
\bold c= ({\rk}, c_1, c_2, c_3)=(2, (0,0), (-4,6), 0 ) \,,
\eeq
where the first Chern class $c_1^r J_r$ should vanish for special unitarity of the structure group, and the second Chern class $c_2^{rs}(X) J_r \wedge J_s$ is represented by another tuple of integers $c_{2, t} \equiv  c_2^{rs} d_{rst} = (-4,6)$ with the following intersection numbers:
\beq\label{inter24}
d_{122}=d_{212}=d_{221}=4,\quad d_{222}=2, \quad\text{all others zero} \,.
\eeq

In this case, an $SU(2)$ bundle $V$ splitting into $\cO_X(a,b) \oplus \cO_X(-a,-b)$, in order to satisfy the Chern vector \eqref{c}, must have $b=-a=1$.
Subsequently, $\mu_B(\cO_X(a,b)) = d_{ijk} (-1,1)^i t^j t^k = 8 t^1 t^2 - 2 (t^2)^2 = 0$ implies that there exists a stability wall along the line
\beq\label{wall-eg1}
\frac{t^2} {t^1}  =4\,, 
\eeq
on which the rank-2 bundle is poly-stable and decomposes into 
\beq \label{decomp_eg}
V \to L_1 \oplus L_2 \,, 
\eeq
with $L_1 = \cO_X(-1,1) $ and $L_2= \cO_X(1,-1)$.
In Ref.~\cite{Anderson:2010ty}, the DT invariant has been obtained from this as 
\beq \label{DTExP9}
DT(X,\bold c = (2, (0,0), (-4,6), 0 ) ; \omega = t^1 J_1 + t^2 J_2 ) = \begin{cases}
-10  {\text{ ,~~if $4< \frac{t^2}{t^1} < \infty$}} \  ,\\ 
~~~~0 {\text{ ,~~if $0<\frac{t^2}{t^1} \leq 4$}} \ , 
\end{cases} 
\eeq 
which is seen to jump when one crosses the stability wall of \eqref{wall-eg1}.

Let us show how this result can be reproduced in the quiver language. 
\begin{figure}[!h]
\centering
\includegraphics[width=7cm]{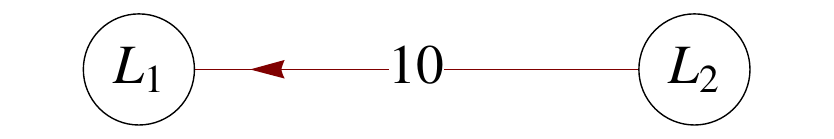}
\parbox{6in}{\caption{\small \it The quiver diagram associated to stable bundles with the fixed Chern class $ ({\rk}, c_1, c_2, c_3)=(2, (0,0), (-4,6), 0 )$ on the $\{2,4\}$ hypersurface in $\IP^1 \times \IP^3$. }\label{fig:ExP9}}
\end{figure}
The quiver associated with the bundle decomposition~\eqref{decomp_eg} is shown in Figure~\ref{fig:ExP9}.
Note that the adjacency matrix is computed using the information about line bundle cohomology on $X$~\cite{cicypackage}:
\bea
\nn
l_{12}&=& h^1(X, L_1 \otimes L_2 ^ *) = h^1(X, \cO_X(-2, 2)) = 10 \ , \\
l_{21}&=&h^1(X, L_2 \otimes L_1^*) = h^1(X, \cO_X(2, -2))=0 \ , 
\eea
where $l_{ij} \equiv h^1(X, L_i \otimes L_j^*)$ are the number of arrows from node $j$ to $i$.
Since $L_1$ and $L_2$ are line bundles, they correspond to $U(1)$ gauge groups and the D-term conditions can be written, using the prescription of \eqref{quiverD}, as 
\begin{eqnarray}
\nn
- \sum\limits_{i=1}^{10}|\phi_{12}^{(i)}|^2 &=&\zeta_1\,,\\ 
\sum\limits_{i=1}^{10}|\phi_{12}^{(i)}|^2&=&\zeta_2\,,  
\end{eqnarray}
where $\phi_{12}^{(i)}$ for $i=1,2,\ldots,10$ are the ten complex fields corresponding to the arrows from node 2 to 1.
This requires as usual that $\zeta_1 + \zeta_2 =0$ and we only have one independent equation.
In case $\zeta_1 <0$ (or equivalently, if the K\"{a}hler moduli sit above the wall), quotienting the D-flat solution space by $U(1)$ gauge transformation gives rise to $\IP^9$ and hence, the DT invariant in this chamber is $-10$. 
On the other hand, if $\zeta_1>0$ (that is, below the wall), there are no solutions at all and hence the DT invariant is $0$.
Note that this is exactly the same wall-crossing as in Eq.~\eqref{DTExP9}.
Indeed, the astute reader would recognize the above D-terms as the GLSM description of $\IP^9$ as a toric variety.

Now, let us compute this in a systematic way by applying Reineke's formula, from which the Poincar\'{e} polynomial of a quiver variety is given by Eq.~\eqref{PtAb} for an Abelian quiver.
In the example at hand, $n=2$ and $\bold d = (1,1)$ has the following three ordered partitions 
\begin{equation}\label{part-eg1}
\bold d^* = ((1,1)),\  ((1,0);(0,1)), \ ((0,1); (1,0)) \ . 
\end{equation}

Let us determine which terms precisely contribute to \eqref{PtAb}.
From \eqref{theta-zeta}, substituting $d_w = 1$ and $\zeta_1 + \zeta_2 = 0$ gives us simply $(\theta_1, \theta_2) = (\zeta_1, \zeta_2) = (\zeta_1, -\zeta_1)$.
Therefore, our stability condition reads as
\begin{equation}\label{muQ-eg1}
\mu_Q(\sum\limits_{l=1}^k \bold{d}^l) = \frac{1}{2} \sum\limits_{v=1}^2 (\sum\limits_{l=1}^k d^l_v) \zeta_v = \frac{\zeta_1}{2} \sum\limits_{l=1}^{k} (d_1^l - d_2^l) > 0 \ ,
\end{equation}
for $k=1, \cdots, s-1$ where $s$ is the number of terms in the three partitions in \eqref{part-eg1}, namely $1,2,2$, respectively.
Note that we use the notation that for the dimension vector, the subscript indexes the node while the superscript indexes the ordering in a particular partition.

Let us start from the $\zeta_1<0$ chamber. 
The first partition $\bold d^* =((1,1))$ has $s=1$ and so \eqref{muQ-eg1} is not applicable as a constraint and hence $((1,1))$ contributes to the Poincar\'e polynomial.
The second partition $\bold d^*=((1,0);(0,1))$ has $s=2$ so \eqref{muQ-eg1} should place one constraint, namely $\frac{\zeta_1}{2} (d^1_1 - d^1_2) > 0$.
However, here $d^1_1 = 1$ and $d^1_2 = 0$, so this constraint is violated since we have chosen $\zeta_1 < 0$. Hence, the second partition does not contribute.
Moving on to the third partition $\bold d^*=((0,1); (1,0))$ which also has $s=2$, the stability constraint requires that 
$\frac{\zeta_1}{2} (d^1_1 - d^1_2) = \frac{\zeta_1}{2} (0 - 1) > 0$, which is obviously satisfied. 
Thus this third partition does contribute.

Therefore, recalling that the only arrows $ v \rightarrow w $ are the 10 from node 2 to 1, and by summing over $\bold d^*$ from the first partition where
$\bold d^1 = (1,1)$ and the third partition where $\bold d^1 = (0,1), \ \bold d^2 = (1,0)$,
Eq.~\eqref{PtAb} now reads, 
\begin{equation}
P(t) = (t^2 - 1)^{-1} \Big[
\underbrace{(-1)^0 t^{\ 2 \cdot 10 \cdot d^1_2 d^1_1 }}_{\bold d^* = ((1,1))} + 
\underbrace{(-1)^1 t^{\ 2 \cdot 10 (d^{l=1}_2 d^{k=1}_1 + d^{l=2}_2 d^{k=1}_1)}}_{\bold d^*=((0,1);(1,0))}
\Big]
= (t^2-1)^{-1} \left[t^{20} - 1\right]\ .
\end{equation}

The situation for the $\zeta_1 > 0$ chamber is similar but the answer is drastically different.
Here, the stability constraint \eqref{muQ-eg1} forces us to consider only partition 1 where $\bold d^1 = (1,1)$ and partition 2 where $\bold d^1 = (1,0), \ \bold d^2 = (0,1)$.
Hence, 
\begin{equation}
P(t) = (t^2 - 1)^{-1} \Big[
\underbrace{(-1)^0 t^{\ 2 \cdot 10 \cdot d^1_2 d^1_1 }}_{\bold d^* = ((1,1))} +
\underbrace{(-1)^1 t^{\ 2 \cdot 10 (d^{l=1}_2 d^{k=1}_1 + d^{l=2}_2 d^{k=1}_1)}}_{\bold d^*=((1,0);(0,1))}
\Big]
= (t^2 - 1)^{-1} \left[ t^{20} - t^{20} \right]
= 0 \ .
\end{equation}

In summary, we have that the Poincar\'e polynomial is
\beq
P(t) =\begin{cases}
\frac{t^{20}-1}{t^2-1} = 1 + t^2 + \cdots + t^{18} \,, ~~\text{if}~ \zeta_1 < 0 \ ,\\ 
~0 ~\;\;\quad\quad\quad\quad\quad\quad\quad\quad\quad{\text{ ,~~if $\zeta_1 > 0$}} \ , 
\end{cases} 
\eeq
which is again consistent with Eq.~\eqref{DTExP9}.


\subsubsection{A Rank-3 Example: $SU(3) \to S[U(1) \times U(1) \times U(1)]$}
On the same Calabi-Yau threefold as in the previous example, 
let us now consider $SU(3)$ bundles, with the fixed total Chern class
\beq
\bold c= ({\rk}, c_1, c_2, c_3)=(3, (0,0), (-12,18), -20 ) \,.
\eeq
In this case, we still have the same stability wall along 
\beq \frac{t^2} {t^1}  =4\,, \eeq
on which the rank-3 bundle is poly-stable and decomposes into 
\beq 
V \to L_1 \oplus L_2 \oplus L_3 \,, 
\eeq
with $L_1 = \cO_X(-2,2) $,  $L_2= \cO_X(1,-1)$ and $L_3 = \cO_X(1,-1)$.
From this fact, again in Ref.~\cite{Anderson:2010ty}, the DT invariant has been obtained as 
\beq \label{DTEx1}
DT(\bold c = (2, (0,0), (-12,18), -20 ) , \omega = t^1 J_1 + t^2 J_2 ) = \begin{cases}
~\,80  {\text{ ,~~if $4< \frac{t^2}{t^1} < \infty$}} \ ,\\ 
~~~0 {\text{ ,~~if $0<\frac{t^2}{t^1} \leq 4$}} \ , 
\end{cases} 
\eeq
which jumps when crossing the wall.

\begin{figure}[!h]
\centering
\includegraphics[width=13cm]{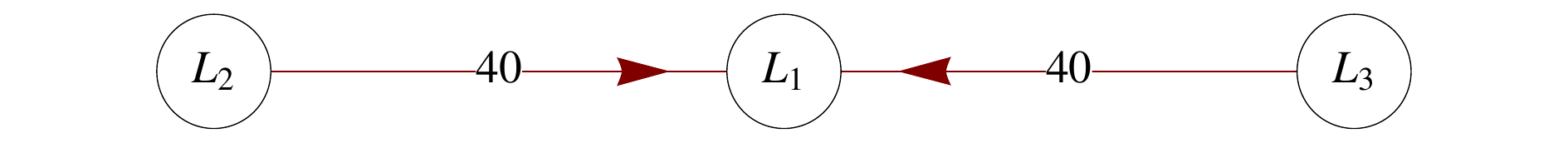}
\parbox{6in}{\caption{\small \it The quiver diagram associated to stable bundles with the fixed Chern class $({\rk}, c_1, c_2, c_3)=(3, (0,0), (-12,18), -20 )$ on the $\{2,4\}$ hypersurface in $\IP^1 \times \IP^3$. }\label{fig:Ex1}}
\end{figure}

The associated quiver is shown in Figure~\ref{fig:Ex1}, 
where the adjacency matrix is computed as
\bea
\nn
&&l_{12}= h^1(X, L_1 \otimes L_2 ^ *) = h^1(X, \cO_X(-3, 3)) = 40 \ , \\
&&l_{13}=h^1(X, L_1 \otimes L_3^*) = h^1(X, \cO_X(-3, 3))=40 \ ,
\eea
with all the other $l_{ij}$ being zero .
Then the D-term conditions are 
\bea
\nn
-|\phi_{12}|^2 - |\phi_{13}|^2 = \zeta_1 \,, \\
\nn
|\phi_{12}|^2 = \zeta_2 \,, \\
|\phi_{13}|^2 = \zeta_3 \,, 
\eea
where $\phi_{12}$ and $\phi_{13}$ are complex forty-dimensional vectors.
The three FI parametres again add up to zero and there remains two independent equations. 
Note that these D-term equations can only be solved if both $\zeta_{2}$ and $\zeta_{3}$ are positive, or equivalently, if the K\"{a}hler moduli sit above the stability wall, ${t^2 \over t^1} = 4$.
In this chamber, quotienting the D-flat solution space by the two $U(1)$ gauge transformations gives rise to $\IP^{39} \times \IP^{39}$ and hence, the DT invariant in this chamber is $80$ (and 0 in the other chamber). 

Of course, one can directly apply Reineke's formula to the quiver in Fig.~\ref{fig:Ex1}, and we find that the same result is readily obtained. 
A word of caution for the details of the calculation is in order, this will be important for our algorithms shortly.
It may seem that we have complete freedom in the choices of the parametres $\zeta_v$ with the only constraint being that they sum to 0, whereby giving us - taking all possible combinations of each being either greater or less than 0 - $2^{n-1}$ chamber divisions.
This is not the case.
We recall from \eqref{bundleD} that the FI-parametres are, up to an overall proportionality constant, simply the mu-slope of the line-bundle summands in the split of the bundle $V$.
The vanishing of this slope, as a function of the K\"ahler moduli $t^{i = 1, \ldots, h^{1,1}(X)}$ defines the stability wall.

The chambers of stability, which thus control supersymmetry and DT invariants, are co-dimension one objects in the K\"ahler cone, which is of dimension $h^{1,1}(X)$ (and not $n-1$).
Therefore, in actual computations, we need to consider either side of the stability wall as determined by mu-slope, giving us the regions of $t^i$ values.
This then in turn gives the signs of the FI parametres $\zeta_v$.
In this above example, we indeed have the Poincar\'e polynomial as
\begin{equation}
P(t) = \left\{
\begin{array}{rcl}
\frac{(t^{80}-1)^2}{(t^2-1)^{2}}, & \mbox{ if } & \zeta_1 < 0, \ 
				    \zeta_{2,3} > 0 \\
0, & \mbox{ otherwise } &  \ . \\
\end{array}
\right.
\end{equation}
That the result of Eq.~\eqref{PtAb} is always a polynomial with non-negative coefficients, as exemplified here,  is highly non-trivial and is guaranteed by Reineke's formalism.

\section{An Algorithmic Outlook}\setall
Having seen our two examples of quiver structure for heterotic moduli on a specific Calabi-Yau threefold in detail, and emboldened by the success of our description, it is immediate that we could look at a multitude of situations.
Indeed, one is naturally led to the classification problem for all possible bundle decompositions: to try to obtain the quiver description for every Chern class 
$
\bold c = ({\rm rk}, c_1, c_2, c_3)
$ given any Calabi-Yau threefold by studying the stability-wall phenomenon, thereby obtaining the splitting behaviour of the bundle moduli as well as the DT invariants. 

This is, of course, a hugely ambitious goal and will involve, in addition to the need to generalizing to non-Abelian summands and quivers with loops, some subtleties which we will shortly discuss in \S\ref{s:ana}.
For now, let us consider the opposite direction.  
That is, we start from the direct sum of line bundles
$
V = \bigoplus\limits_{i=1}^n L_i \,, 
$
where all the summands have their slops vanishing on a common wall inside the K\"ahler cone.
As an upside, this procedure is systematic enough and can easily be put into a computer code.
Unfortunately, one can only get partial information about DT invariants since the bundle decompositions sometimes include non-Abelian pieces. 
Although non-Abelian sub-bundles can also be described by a quiver, they are not completely determined by Chern classes, unlike line bundles.
Here, for the purpose of introducing quiver structure, we content ourselves with this opposite direction.

\subsection{Some Analytics}\label{s:ana}
Before we move on to the database of threefolds, let us see what generalities one could draw from Eq.~\eqref{PtAb}.
The sole input to Reineke's formula is the quiver adjacency matrix $l_{ij}$, which is determined by the bundle information.
Here, we need to point out a subtlety.
In our case of complete split into line bundles for a bundle with given Chern vector, we must be careful of contributions coming from sheaves.
In this case, such torsion sheaves would be supported on sub-varieties of $X$ of co-dimension at least two since line-bundles are in one-one correspondence with divisors which are of co-dimension one.
In other words, for any fixed Chern vector, contributions from both line bundles and sheaves will be present.

Of course, so long as the $\ext^1$-group is unaffected by the torsion, the adjacency matrix, and subsequently any details of the quiver variety and bundle moduli, will be unaffected.
In this case, assuming that we are only dealing with line-bundle splits is sufficient.
In \cite{LiQin, LiQin2}, it was shown that this is the case for our example of the threefold in Eq.~\eqref{X24}.
In general, we need to be careful.

Nevertheless, we can conclude many analytic results with specific forms of the input.
Suppose we have a rank-2 bundle $V = L_1 \oplus L_2$, then the most general form of the corresponding loop-less quiver is
\begin{equation}
l^{n = 2} = \left(
\begin{matrix}
0 & l_{12} = \ext^1(L_2, L_1) \\
0 & 0
\end{matrix}
\right) \ ,
\end{equation}
where $L_i$ are rank-1 sheaves.
Note that the diagonal vanishes to avoid self-adjoining loops and there can only be one off-diagonal (which we have chosen, without loss of generality, to be upper-right) for otherwise there will be bi-directional arrows (loop of length 2) between nodes 1 and 2.
There is only one independent FI parametre, $\zeta_1$, and the resulting Poincar\'e polynomial is
\begin{equation}
P(t) = \left\{
\begin{array}{rcl}
(t^{2 l_{12}}-1)(t^2-1)^{-1}, & \mbox{ if } & \zeta_1 < 0 \\
0, & \mbox{ if } & \zeta_1 > 0 \ . \\
\end{array}
\right.
\end{equation}

Moving on to the general rank-3 case, the situation becomes more unwieldy; this is because unlike the $n=2$ situation where there is only one free FI parametre, as discussed above, we need to determine the chambers in the K\"ahler cone, which requires information on the mu-slope and the precise split of the line bundle summands.
For this, we will need to know the topology of the Calabi-Yau threefold $X$.
Suppose we have a rank-3 split $V = L_1 \oplus L_2 \oplus L_3$. Up to re-ordering of the nodes, the adjacency matrix without loops can take the form of $l^{n=3} = {\tiny \left(
\begin{matrix}
0 &l_{12}& l_{13} \\
0 & 0    & l_{23} \\
0 & 0    & 0
\end{matrix}
\right)}$, with the entries depending on the $L_i$, via the appropriate bundle cohomologies on $X$.
Let us assume that the three entries $l_{12}, l_{13}, l_{23}$ are all nonzero and that the integral two-vectors $c_1(L_2)$ and $c_1(L_3)$ are proportional to each other by a positive scalar constant. 
In this case, the Poincar\'e polynomial takes the form
\begin{equation}
P(t) = \left\{
\begin{array}{rcl}
\frac{(t^{2(l_{13} + l_{23})}-1)(t^{2l_{12}}-1)}{(t^2-1)^{2}}, & \mbox{ if } & \zeta_1 < 0, \ 
				    \zeta_{3} > 0 \\
0, & \mbox{ otherwise } &  \ . \\
\end{array}
\right.
\end{equation}
We can go on to analyze the Poincar\'e polynomial for more general rank-3 cases or for higher ranks.
However, perhaps fatigued by abstraction, let us move on to concreteness.

\subsection{CICYs: a Plenitude of Examples}\label{s:CICY}
Before we start scanning the bundle decompositions, a Calabi-Yau threefold has to be specified. 
Since $h^{1,1}(X) >1$ is required for nontrivial wall-crossing phenomena, we demand $h^{1,1}(X)=2$ as for the minimal case.
A dataset instantly springs to mind, this is the famous and oldest database of Calabi-Yau threefolds \cite{Candelas:1987kf,Gagnon:1994ek,cicypackage}, the so-called CICYs, or complete intersections in products of projective spaces.
From this list over which line bundle cohomologies are rather easily computed, 36 manifolds are found to have $h^{1,1}=2$. 

It is not immediately conducive to exhaustively study all 36 manifolds, so, for convenience, let us demonstrate with the following four examples:
\beq \nn
X_1=\left[
\begin{array}
[c]{c}%
\IP^1\\
\IP^4
\end{array}
\left|
\begin{array}
[c]{cc}%
0 & 2\\
3 &2 
\end{array}
\right.  \right]; ~~
X_2=\left[
\begin{array}
[c]{c}%
\IP^2\\
\IP^3
\end{array}
\left|
\begin{array}
[c]{cc}%
2 & 1\\
1 & 3
\end{array}
\right.  \right];~~
X_3=\left[
\begin{array}
[c]{c}%
\IP^2\\
\IP^2
\end{array}
\left|
\begin{array}
[c]{c}%
3\\
3 
\end{array}
\right.  \right];~~
X_4=\left[
\begin{array}
[c]{c}%
\IP^1\\
\IP^3
\end{array}
\left|
\begin{array}
[c]{c}%
2\\
4 
\end{array}
\right.  \right] \,,
\eeq
the last two being the only hypersurface cases amongst the 36.

On each of the four Calabi-Yau threefolds $X_i$ ($i=1,2,3,4$), we can construct rank-$n$ special unitary bundles $V$ for $n=2, 3, 4, 5$, which are the cases of phenomenological interest.
To make the scan finite, we will need to set some artificial bounds on the Chern class entries of the summands in the splitting of $V$. 
Furthermore, as aforementioned, we only deal with those quivers which do not involve loops.
Finally, for convenience, we will adhere to only bona fide line bundles for the summands, instead of general rank-1 sheaves.

In summary, we will require, for $V = \bigoplus\limits_{i=1}^n L_i$, 
the following:
\begin{enumerate}
\item $SU(n)$: the bundle $V$ is special unitary, i.e., $\sum\limits_{i=1}^n c_1(L_i) = 0 \in \IZ^2$ ;

\item Bound: summands are line bundles $\cO_X(a,b)$ whose entries $a,b$ are bounded by $\pm 3$, that is, $c_1(L_i) \in [-3, 3]^2$ ;

\item Stability: all the summands $L_i$ have a vanishing slope on a common wall in the K\"ahler cone, and hence, there exists a nontrivial vector $(t^1, t^2)$ in the positive quadrant such that $c_1^r(L_i) t^s t^u d_{rsu} =0$, for all $i= 1, \dots, n$ ; 

\item Loop-less: the associated quiver has no directed loops. In other words, the line bundle cohomologies $l_{ij} = h^1(X, L_i \otimes L_j^*)$ are such that there is no closed path $i_1 \to i_2 \to \cdots \to i_m \to i_1$ of length $m \geq 1$, with the linking numbers $l_{i_1 i_2}, l_{i_2 i_3}, \dots, l_{i_m i_1}$ all positive along the path.
\end{enumerate}

How many such bundles are there on our four manifolds?
The statistics for this scan is summarised in Table~\ref{t:stat}.
We see that even with our constraints and bound on Chern classes there is plenty of examples to analyze.

\renewcommand{\arraystretch}{1.5}
\begin{table}[h!t!]
\begin{center}
\beq\nn
\ba{|c|c|c|c|c|c|}\hline
\mbox{CICY}~X &  ~n=2~ & ~n=3~ & ~n=4~ &
~n=5~ \\ \hline \hline
&&&&\\[-6mm]
~~~X_1=\scriptsize
{\left[
\begin{array}
[c]{c}%
\IP^1\\
\IP^4
\end{array}
\left|
\begin{array}
[c]{cc}%
0 & 2\\
3 &2 
\end{array}
\right.  \right]}~~~
&
3
& 3
& 6
& 6
\\[3mm]\hline
&&&&\\[-6mm]
X_2=\scriptsize\left[
\begin{array}
[c]{c}%
\IP^2\\
\IP^3
\end{array}
\left|
\begin{array}
[c]{cc}%
2 & 1\\
1 & 3
\end{array}
\right.  \right]
& 9
& 12
& 28
&  41
\\[3mm] \hline
&&&&\\[-6mm]
X_3=\scriptsize\left[
\begin{array}
[c]{c}%
\IP^2\\
\IP^2
\end{array}
\left|
\begin{array}
[c]{c}%
3\\
3 
\end{array}
\right.  \right]
& 9
& 13
& 29
& 43
\\[3mm] \hline
&&&&\\[-6mm]
X_4=\scriptsize\left[
\begin{array}
[c]{c}%
\IP^1\\
\IP^3
\end{array}
\left|
\begin{array}
[c]{c}%
2\\
4 
\end{array}
\right.  \right] 
& 7
& 11
& 25
& 39
\\[3mm] \hline
\ea\nn
\eeq
\caption{\small \it Statistics for the number of possible decompositions of $SU(n)$ bundles into the sum of $n$ line bundles, $V = \bigoplus\limits_{i=1}^n L_i$ ($n=2,3,4,5$), on the four Calabi-Yau threefolds $X_{1,2,3,4}$ under our 4 constraints. The line bundles $L_i$ were constrained so that $c_1(L_i) \in [-3,3]^2$, $\sum\limits_{i=1}^n c_1(L_i) = 0$ and all the slopes $\mu_B(L_i)$ can vanish at a common locus in the K\"ahler cone. Furthermore, only those decompositions whose  associated quivers do not involve loops were selected from the scan.}
\label{t:stat}
\end{center}
\end{table}
\begin{figure}[!h!t]
\centering
$\begin{array}{l}
\includegraphics[width=8cm]{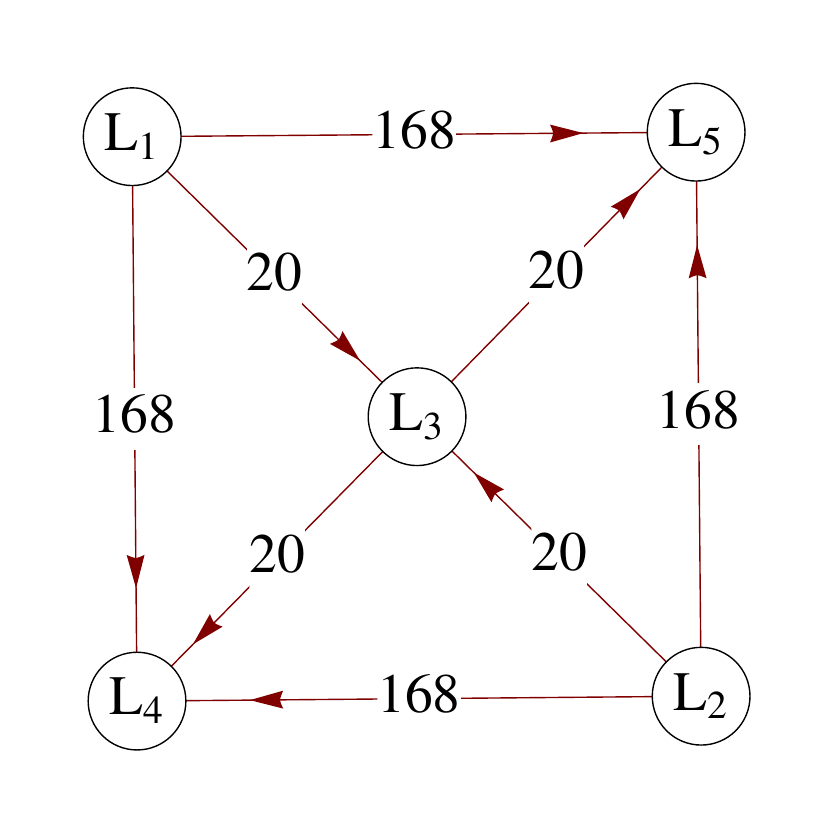}\end{array}$
$l_{ij} = \left(
\begin{array}{ccccc}
 0 & 0 & 0 & 0 & 0 \\
 0 & 0 & 0 & 0 & 0 \\
 20 & 20 & 0 & 0 & 0 \\
 168 & 168 & 20 & 0 & 0 \\
 168 & 168 & 20 & 0 & 0 \\
\end{array}
\right)$
\parbox{6in}{\caption{\small \it An example of rank-$5$ quivers on $\{2,4\}$ hypersurface $X_4$ in $\IP^1 \times \IP^3$. On the stability wall, this quiver describes the split of vector bundles into direct sum of the five line bundles $L_{1,2} = \cO_{X_4} (3,-2)\;,~ L_3=\cO_{X_4}\;,~L_{4,5} = \cO_{X_4} (-3,2)$. 
}\label{fig:Ex3}}
\end{figure}

As an illustration of the large resulting list, let us consider the following rank-$5$ example on $X_4$:  
\beq
V = \bigoplus\limits_{i=1}^5 L_i\;, ~\mbox{with}~ 
L_{1,2} = \cO_{X_4} (3,-2)\;,~ L_3=\cO_{X_4}\;,~L_{4,5} = \cO_{X_4} (-3,2) \ .
\eeq  
This is a rather non-trivial case shown in Figure \ref{fig:Ex3};
the adjacency matrix is also given for reference.
We can then apply the formula \eqref{PtAb} to readily find that
\begin{equation}
P(t) = \left\{
\begin{array}{rcl}
\frac
{
t^{1504}-t^{1424}-4 t^{1128}+4 t^{1088}+4 t^{416}-4 t^{376}-t^{80}+1
}
{
\left(t^2-1\right)^4
},
& \mbox{ if } &
\zeta_{1,2} > 0, \ \zeta_{4,5} < 0; \\
0,  & & \mbox{otherwise } \ .
\end{array}
\right.
\end{equation}
Again, it is remarkable that this is a polynomial with non-negative coefficients only.

\subsection{Division Pattern of the K\"ahler Cone}\label{s:curve}
In the examples of \S\ref{s:eg}, we have seen that the K\"ahler cone (that is, the positive quadrant in $\IR^2$) is divided into two chambers, with a single stability wall being the flat line, $\frac{t^2}{t^1} = 4$, between the two.  
In general, however, one would naturally expect that the situation is much more intricate.
Firstly, the walls of marginal stability could be curved; for instance, in pure $SU(2)$ Seiberg-Witten theory, which provides a prototypical example of wall-crossing in field theory, the two-dimensional moduli space for the Coulomb vevs is divided into two chambers and the stability wall is indeed curved with a circular topology (cf.~\cite{Pioline:2011gf}).
Furthermore, one may in principle expect the presence of multiple walls and the K\"ahler cone is divided into many regions separated by non-linear walls. 
In this subsection, we will see examples of such patterns of the K\"ahler cone.  

It turns out that, on the Calabi-Yau threefolds with $h^{1,1}=2$, constructed either as CICYs~\cite{Candelas:1987kf,Gagnon:1994ek,cicypackage} or as toric hypersurfaces~\cite{Kreuzer:2000qv, Kreuzer:2000xy}, such behaviours can never occur for our Abelian splits.  
However, by relaxing either of the two constraints (that is, the Abelian property or the $h^{1,1}=2$ condition), one can easily obtain some examples with interesting division pattern of the K\"ahler cone. 
Indeed, an illustrative example of multiple-wall structure has been given for non-Abelian splits~\cite{Anderson:2010tc}. Here, we present another interesting one, in which the stability wall is a curved locus. 

For this, one needs to go higher than $h^{1,1}=2$. 
Let us consider the following Calabi-Yau three-fold 
\beq
X=\left[
\begin{array}{c|c}
\IP^1 & 2\\[-0.4cm]
\IP^1 & 2\\[-0.4cm]
\IP^2 & 3
\end{array}
\right] \,,
\eeq
which is a hypersurface of multi-degree $(2,2,3)$ in $\IP^1 \times \IP^1 \times \IP^2$.
The K\"ahler cone for this threefold is simply the positive octant, $t^r >0$, $r=1,2,3$, and the triple intersection numbers are encoded into the coefficients of the polynomial
\begin{equation} \label{int223}
d(x_1,x_2,x_3) = 18 x_1 x_2 x_3 + 6 x_1 x_3^2 + 6 x_2 x_3^2 \ .
\end{equation}
Let us fix the total Chern class as
\beq
\bold c = ( {\rm rk} , c_1, c_2, c_3)= (2, (0,0,0), (-8,4,6), 0)
\eeq
and consider $SU(2)$ bundles. 
One then finds the split is 
$\cO_X(-1,1,1) \oplus \cO_X(1,-1,-1)$.
\begin{figure}[!h!t]
\centering
\includegraphics[width=7cm]{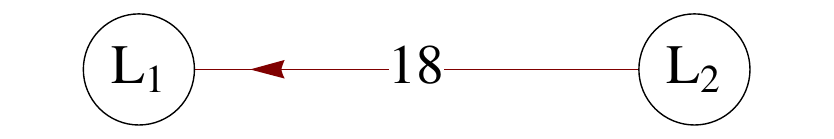}
\parbox{6in}{\caption{\small \it The quiver diagram associated to stable bundles with the fixed Chern class $({\rk}, c_1, c_2, c_3)=(2, (0,0,0), (-8,4,6), 0)$ on the $\{2,2,3\}$ hypersurface in $\IP^1 \times \IP^1 \times \IP^2$. 
}\label{fig:3dQuiver}}
\end{figure}
\begin{figure}[!h!t]
\centering
\includegraphics[width=8cm]{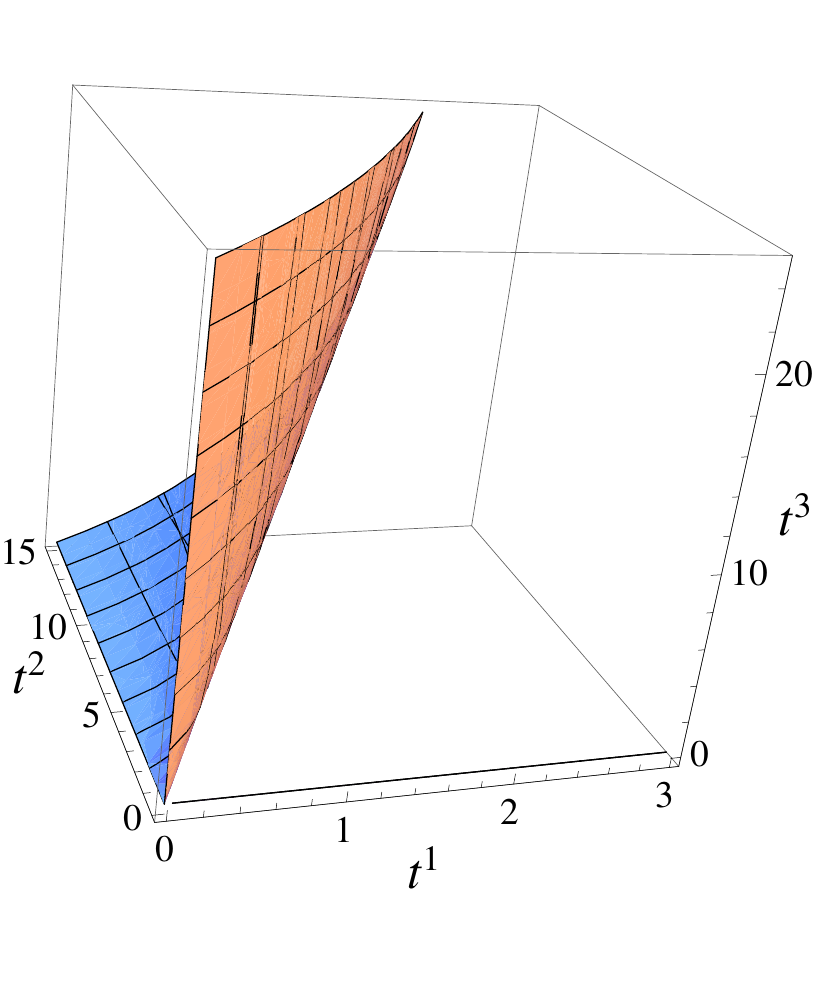}
\parbox{6in}{\caption{\small \it The stability wall, defined by $6 t^1 t^2 + 10 t^1 t^3 - 2 t^2 t^3 = 0$ inside the positive octant, on which the $SU(2)$ bundle decomposes into the sum of two line bundles. The quiver in Figure~\ref{fig:3dQuiver} describes the split as well as the moduli space of stable bundles near this wall.  
}\label{fig:3dWall}}
\end{figure}
Thus, using the intersection numbers~\eqref{int223}, we see that $\mu_B(\cO_X(-1,1,1)) = 0$ implies that there is a stability wall located along the curved locus
\beq \label{3dWall}
6 t^1 t^2 + 10 t^1 t^3 - 2 t^2 t^3 = 0 \ ,
\eeq
whereupon the rank-$2$ bundle becomes poly-stable and decomposes into 
\beq
V \to L_1 \oplus L_2 \,,
\eeq
with $L_1= \cO_X(-1,1,1)$ and $L_2 = \cO_X(1,-1,-1)$.
The associated quiver is depicted in Figure~\ref{fig:3dQuiver} and the wall inside the K\"ahler cone, which is seen to be curved, is shown in Figure~\ref{fig:3dWall}.

\vspace{0.5in}

\centerline{ \reflectbox{\ding{167}}---\ding{69}---\ding{167} }

\section*{Acknowledgments}
Y.-H.~H gratefully acknowledges Eric Sharpe for discussions, and
would like to thank the Science and Technology Facilities Council, UK, for an Advanced Fellowship and grant ST/J00037X/1, the Chinese Ministry of Education, for a Chang-Jiang Chair Professorship at NanKai University, US NSF grant CCF-1048082, as well as City University, London and Merton College, Oxford, for their enduring support.

\noindent
S.-J.~L would like to thank Kang-Sin Choi, Hirotaka Hayashi, Yoon Pyo Hong, Andre Lukas, Zhaolong Wang and Piljin Yi for helpful discussions. 
He is particularly indebted to Yoon Pyo Hong and Piljin Yi for motivating and stimulating discussions during the early stage of the work. 
He also thanks Lara Anderson for kindly pointing out and explaining related earlier works. The work of S.-J.~L was supported in part by the National Research Foundation of Korea (NRF) under grant number 2010-0013526.


\end{document}